\documentclass[12pt]{article}
\usepackage{latexsym, epsfig, graphics, amsmath}
\newcommand{\be}{\begin{equation}}
\newcommand{\ee}{\end{equation}}
\newcommand{\bq}{\begin{eqnarray}}
\newcommand{\eq}{\end{eqnarray}}
\newcommand{\cue}{{\bf q}}
\newcommand{\dbsm}{\sum_{\sigma_{1}<\sigma}\sum_{\sigma<\sigma_{2}}}
\newcommand{\sqsm}{\sqrt{(\sigma-\sigma_{1})(\sigma_{2}-\sigma_{1})
(\sigma_{2}-\sigma)}}
\newcommand{\wss}{W(\sigma_{1},\sigma_{2})}
\newcommand{\ssm}{\frac{\sigma_{2}-\sigma_{1}}{(\sigma_{2}-\sigma)
(\sigma -\sigma_{1})}}
\newcommand{\bfv}{{\bf v}}
\newcommand{\qsm}{(\sigma -\sigma_{1}) (\sigma_{2} -\sigma_{1})
(\sigma_{2} -\sigma)}
\begin{document}
\begin{titlepage}
\today          \hfill 
\begin{center}

\vskip .5in

{\large \bf Scalar Field Theories On The World Sheet: Cutoff Independent
Treatment }\footnote{This work was supported by the Director, Office of
Science, Office of High Energy Physics of the U.S. Department of Energy
under Contract DE-AC02-05CH11231.}

\vskip .50in


\vskip .5in
Korkut Bardakci \footnote{Email: kbardakci@lbl.gov}

{\em Department of Physics\\
University of California at Berkeley\\
   and\\
 Theoretical Physics Group\\
    Lawrence Berkeley National Laboratory\\
      University of California\\
    Berkeley, California 94720}
\end{center}

\vskip .5in

\begin{abstract}

Following earlier work on the same topic, we consider once more scalar
field theories on the world sheet parametrized by the light cone
coordinates. For most of the way, we use the same approach as in the
previous work, but there is an important new development. To avoid the
light cone singularity at $p^{+}=0$, one world sheet coordinate had to be
 discretized, introducing a cutoff into the model. In the earlier work, this
cutoff could not be removed, making the model unreliable. In the present
article, we show that, by a careful choice of the mass counter term,
both the infrared singularity at  $p^{+}=0$ and the ultraviolet mass
divergences can be simultaneously eliminated. We therefore finally have 
a cutoff independent model on a continuously parametrized world sheet.
We study this model in the mean field approximation, and as before,
we find solitonic solutions. Quantizing the solitonic collective
coordinates gives rise to a string like model. However, in contrast
to the standard string model, the trajectories here are not
in general linear but curved.

\end{abstract}
\end{titlepage}

\newpage
\renewcommand{\thepage}{\arabic{page}}
\setcounter{page}{1}
\noindent{\bf 1. Introduction}
\vskip 9pt

The present article is likely the last in a  series of several
articles, starting with [1]. The idea behind this work was to sum the 
planar graphs of field theory on a world sheet parametrized by the
light cone variables, based on 't Hooft's pioneering paper [2]. The
original field theory studied in this approach was a scalar with
$\phi^{3}$ interaction, and this was later generalized to more
complicated and more interesting models [3, 4]. This paper is a follow up to
 reference [5], using mostly the same approach developed in previous
publications. The models under consideration are scalar
theories with both only $\phi^{3}$ interaction and also with a mixture
of $\phi^{4}$ added. In [5], using both mean field and variational
approximations on the world sheet, solitonic classical solutions
on the world sheet were
constructed, and a certain set of quantum fluctuations about the
solitonic solutions were shown to have a string like spectrum.

These computations  suffer from two kinds of
divergences: One of them is the  field theoretic ultraviolet
divergences, which are eliminated by the standard renormalization
procedure.
  The second one is an 
 infrared divergence due to the choice of the light cone
coordinates. In the previous work, this infrared problem was temporarily
 circumvented by the discretizing the $\sigma$ coordinate on the world sheet
in steps of $a$, but then, several quantities of physical interest, such as
the ground state energy, diverged as $1/a^{2}$ in the limit
$a\rightarrow 0$. This casts serious doubt on the whole approach;
in a massive theory that we are dealing with,
 since there is no infrared problem, this singularity should be spurious [6].
This also prevents us from discussing the continuum limit on the world sheet,
which is after all the model of interest.

The main result of this paper is that this singularity is indeed spurious,
and it can be eliminated by a mass counter term. It is surprising and
highly satisfying that the same counter terms that are needed to cancel
the ultraviolet mass divergences  also automatically cancel the infrared
singularity at $a=0$, a possibility which was overlooked 
  in [5]. As we shall see later in section 4,
 the key to the elimination of this singularity
is in taking the  limit $a\rightarrow 0$ with proper care.

With the singularity at $a=0$ eliminated, there is no obstacle to taking the
continuum limit, and to applying the mean field approximation to the
continuum model. Just as in [5], we construct the solitonic solution and 
 with the soliton as background, we  study the
quantum fluctuations.
The solitonic solutions are of interest because they describe a 
non-perturbative feature of field theory. Also, as we shall see later,
the soliton emerges from  the summation of a dense set of graphs
on the world sheet, which can be thought of as the condensation of
these graphs. The existence of such a condensate on the world sheet
is naturally expected to lead to a string description, an old idea
that motivated some of the early work on this subject [7, 8].

To make the present work self contained and comprehensible, sections
2 and 3, as well as the early part of 4, are mostly devoted to 
the review of the earlier work, especially of reference [5].
In section 2,   the world sheet picture of the planar graphs
of the  $\phi^{3}$ field theory is reviewed,
 and in section 3, we describe
the world sheet field theory, developed in [9], which reproduces these
graphs. This theory is constructed in terms of a complex scalar field 
 and a two component fermion field. In its formulation,
 a central role is played by the field
$\rho$ (eq.(4)), a composite of the fermions, which measures the density
of the graphs on the world sheet.

In section 4, we discuss in detail the construction of the classical
solitonic solutions of the world sheet field equations. These solutions are 
hybrid in character: The complex scalar $\phi$ is treated classically, but
the fermions  are still operators that satisfy the
usual anti commutation relations. The reason for this hybridization
 is that there are certain
overlap relations between interaction vertices (eqs. (20) and (21)),
similar to those encountered in string theory, which we would like to
treat exactly at this stage of the computation. These overlap relations are
crucial for the successful renormalization of the model and
 the elimination of the
 singularity at  $a=0$. They follow from the algebra satisfied by the bilinears
of the fermions (eq.(19)), and to keep this algebra intact,
we solve the equations of motion  by means of a ``hybrid''
 ansatz (eq.(23)). This ansatz is then used to
compute the corresponding Hamiltonian (eq.(25)), and the result is simplified
as much as possible with the help of the overlap relations.
 We show that in this final  form of the Hamiltonian,
the ultraviolet mass divergences  that appear in dimensions $D=2, 4$
can be canceled by a mass counter term.
 
So far, the sigma coordinate is still discretized; in fact, without
 such a discretization, the overlap relations would be difficult to
write down without encountering singular expressions. In section 5, we
show that after scaling the scalar and fermionic fields by $\sqrt{a}$,
the continuum limit $a\rightarrow 0$ can be taken directly without any
difficulty. The result (eq.(30)) is free of both infrared and ultraviolet
singularities, except for a log singularity in the coupling constant at
$D=4$. This singularity can also be circumvented by coupling constant
renormalization. It is remarkable that the same mass term that cancels
the ultraviolet divergence also eliminates the infrared singularity.

 The continuum limit comes with an additional bonus: The model
is now invariant under the subgroup of Lorentz transformations that preserve
the light cone, including the boost $K_{1}$ along the special direction 1.
Invariance under this boost, broken when the sigma coordinate is discretized,
 is restored in the continuum limit.
We hasten to add that, a priory, there is nothing wrong with this
discretization, which amounts to the compactification of the light cone
coordinate $x^{-}$ [10, 11]. It is just that one can now decompactify without
encountering any obstacles.

Now that we have a satisfactory model, free of divergences, we  try
approximations to make it tractable. In section 6, we introduce the
mean field method, which was frequently used in the earlier work.
In this approximation, the fields are replaced by their ground state
expectation values, which are assumed to be independent of the
coordinates $\sigma$ and $\tau$. This strategy is familiar from field theory.
To find the ground state of the model in this approximation, we have to solve
the classical equations of motion (eq.(36)). Although these equations
can be investigated in all generality, in order to get a simple result
in closed form, we consider only the limit of large $\rho$, or a dense
set of graphs on the world sheet. As we have mentioned earlier, this limit is
interesting from the point of view of solitonic solutions and string
formation. In transverse dimensions $D=1$ and $D=2$, we get lower bounds
on the coupling constants (eqs.(40)), necessary for the formation
of the solitonic condensate on the world sheet. It is possible that
the critical values of the coupling constants mark a phase transition
from the weak coupling perturbative phase to the strong coupling 
condensate phase.
 We also compute the masses
of the ground stets in various dimensions, and at least in this
approximation, there is no tachyonic instabilities.

In section 7, we consider an interaction which is an admixture of  
$\phi^{3}$ and $\phi^{4}$ for $D=1$, to see whether the simultaneous
cancellation of the ultraviolet and the infrared singularities 
by a mass counter term still goes through. It is highly satisfactory
that this cancellation still works as before, showing that it not
a special feature of a pure $\phi^{3}$ interaction.

Having constructed the solitonic solutions, in section 8, we study the
 quantum fluctuations in the solitonic background. Here, we  focus
exclusively on a particular set of fluctuations, which were also studied
in the previous work [5]. They come about because the soliton, having
a definite location, breaks the translation symmetry of the model (eq.(17)).
It is then the standard procedure to introduce collective coordinates
corresponding to translations. Upon quantization, these collective modes
restore the spontaneously broken translation symmetry. They can therefore
be identified  as the Goldstone modes, and are expected to dominate
the low energy regime. Also, as in the earlier work, we expect them
to play an
important role in string formation on the world sheet. As an additional bonus,
 in the mean field approximation, their contribution can be computed
 in closed form (eqs.(52) and (59)), and the result is a generalized free
 field theory.

 In section 9, we determine the spectrum of the model, and find that,
unlike the conventional string theory,
the Regge trajectories are no longer linear. Only in the asymptotic
limit when the density of graphs, $\rho$, tends to infinity, the
standard string model with linear trajectories is recovered. 

We end this section by comparing the present paper with the previous
work, especially [5]. The main new result is that we can
let the lattice spacing $a$ go to zero, without encountering any
singularities. This makes it possible to study the model on a
continuous world sheet. In contrast, in the previous work, strictly
speaking, the limit
$a\rightarrow 0$ did not exist.
 We also have some new technical results, such as
masses of the ground state excitations, and a new string like model
with curved trajectories. But the main advance is in reliability;
finally, there is a treatment of the scalar field theory on a world
sheet free of cutoffs. Going back to the first sentence of this section,
now that the model is  free of obvious problems, we feel that the
original project has reached a natural end.
 Of course, this remark only
applies to scalar theories and the light cone setup. More physical
theories with spin and Lorentz invariance [3, 4, 12, 13]
 remain as important problems for
future research.

\vskip 9pt
\noindent{\bf 2. The World Sheet Picture}
\vskip 9pt

The planar graphs of $\phi^{3}$ can be represented [2] on a world sheet
parametrized by the light cone coordinates $\tau=x^{+}$ and
$\sigma=p^{+}$ as a collection of horizontal solid lines (Fig.1), where
the n'th line carries a D dimensional transverse momentum $\cue_{n}$.
\begin{figure}[t]
\centerline{\epsfig{file=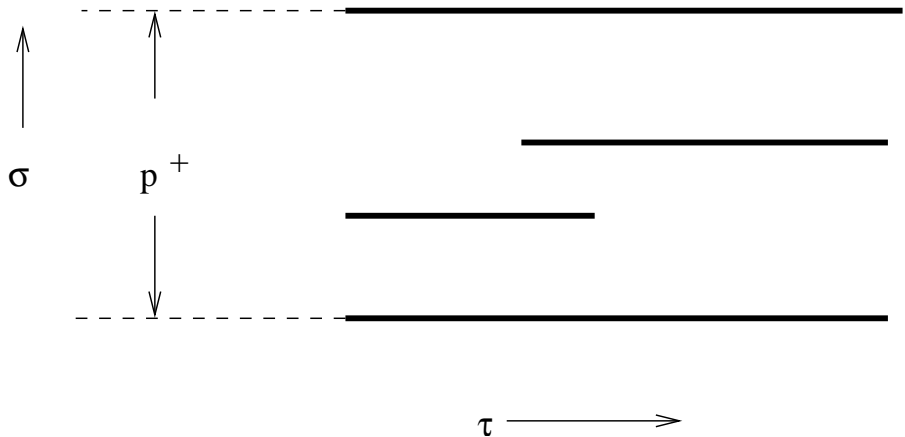, width=12cm}}
\caption{A Typical Graph}
\end{figure}
Two adjacent solid lines labeled by n and n+1 correspond to the light
cone propagator
\be
\Delta({\bf p}_{n})=\frac{\theta(\tau)}{2 p^{+}}\,\exp\left(
-i \tau\, \frac{{\bf p}_{n}^{2}+ m^{2}}{2 p^{+}}\right),
\ee
where ${\bf p}_{n}= \cue_{n}-\cue_{n+1}$ is the momentum flowing through
the propagator. A factor of the coupling constant g is inserted
 at the beginning and at the end of each line, where the interaction
takes place. Ultimately, one has to integrate over all possible
locations and lengths of the solid lines, as well as over the
momenta they carry.

The propagator (1) is singular at $p^{+}=0$. It is well known that 
this is a spurious singularity peculiar to the light cone picture.
To avoid this singularity  temporarily,
it is convenient to
discretize the $\sigma$ coordinate in steps of length $a$.
 A useful way of visualizing the discretized world sheet is
pictured in Fig.2. The boundaries of the propagators are marked by
solid lines as before, and the bulk is filled by dotted lines spaced
at a distance $a$.
For convenience, the $\sigma$ coordinate
 is compactified by imposing periodic
boundary conditions at $\sigma=0$ and $\sigma=p^{+}$. In contrast, the
boundary conditions at $\tau=\pm \infty$ are left arbitrary. We stress that,
 in a massive and hence infrared finite theory, the $p^{+}=0$
 singularity should be absent. We will show how to eliminate
 it in the following sections, and this will allow us to go from a discrete
to a continuous world sheet.

\begin{figure}[t]
\centerline{\epsfig{file=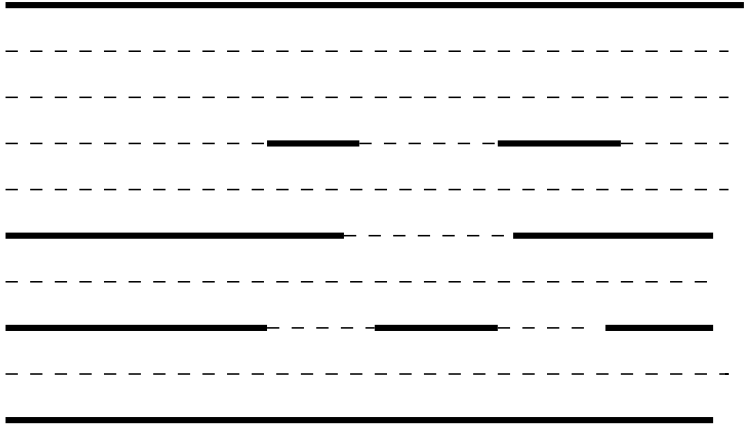, width=10cm}}
\caption{Solid And Dotted Lines}
\end{figure}

\vskip 9pt

\noindent{\bf 3. The World Sheet Field Theory}

\vskip 9pt

It was shown in [9] that the light cone graphs described above are
reproduced by a world sheet field theory, which we now briefly review.
We introduce the complex scalar field $\phi(\sigma,\tau,\cue)$ and
its conjugate $\phi^{\dagger}$, which at time $\tau$
 annihilate (create) a solid line with coordinate $\sigma$ carrying
momentum $\cue$. They satisfy the usual commutation relations
\be
[\phi(\sigma,\tau,\cue),\phi^{\dagger}(\sigma',\tau,\cue')]=
\delta_{\sigma,\sigma'}\,\delta(\cue-\cue').
\ee
The vacuum, annihilated by the $\phi$'s, represents the empty world sheet.

In addition, we introduce a two component fermion field $\psi_{i}(
\sigma,\tau)$, $i=1,2$, and its adjoint $\bar{\psi}_{i}$, which
satisfy the standard anti commutation relations. The fermion with
$i=1$ is associated with the dotted lines and $i=2$ with the solid
lines. The fermions are needed to avoid unwanted configurations
on the world sheet. For example, multiple solid lines generated by
the repeated application of $\phi^{\dagger}$ at the same $\sigma$
would lead to over counting of the graphs. These redundant states can
be eliminated by imposing the constraint
\be
\int d\cue\, \phi^{\dagger}(\sigma,\tau,\cue)\phi(\sigma,\tau,\cue)
=\rho(\sigma,\tau),
\ee
where
\be
\rho=\bar{\psi}_{2}\psi_{2},
\ee
which is equal to one on solid lines and zero on dotted lines. This
constraint ensures that there is at most one solid line at each
site.

Fermions are also needed to avoid another set of unwanted configurations.
Propagators are assigned only to adjacent solid lines and not to
non-adjacent ones. To enforce this condition, it is convenient to
define, 
\be
\mathcal{E}(\sigma_{i},\sigma_{j})=\prod_{k=i+1}^{k=j-1}\left(
1-\rho(\sigma_{k})\right),
\ee
for $\sigma_{j}>\sigma_{i}$, and zero for $\sigma_{j}<\sigma_{i}$.
The crucial property of this function is that it acts as a projection: 
It is equal to one when the two lines at $\sigma_{i}$ and $\sigma_{j}$
are separated only by the dotted lines; otherwise, it is zero. With the
help of $\mathcal{E}$, the free Hamiltonian can be written as
\bq
H_{0}&=&\frac{1}{2}
\sum_{\sigma,\sigma'}\int d\cue \int d\cue'\,\frac{\mathcal
{E}(\sigma,\sigma')}{\sigma'-\sigma} \left((\cue-\cue')^{2}+ m^{2}
\right)\nonumber\\
&\times& \phi^{\dagger}(\sigma,\cue) \phi(\sigma,\cue)
 \phi^{\dagger}(\sigma',\cue') \phi(\sigma',\cue')\nonumber\\
&+&\sum_{\sigma} \lambda(\sigma)\left(\int d\cue\,
 \phi^{\dagger}(\sigma,\cue) \phi(\sigma,\cue) -\rho(\sigma)\right),
\eq
where $\lambda$ is a Lagrange multiplier enforcing the constraint (3).
The evolution operator $\exp(-i \tau H_{0})$, applied to states,
generates a collection of free propagators, without, however, the
prefactor $1/(2 p^{+})$.

One can also think of 
the Lagrange multiplier $\lambda(\sigma, \tau)$ as an Abelian
gauge field on the world sheet. The corresponding gauge
transformations are [14]
\bq
\psi &\rightarrow& \exp\left(- \frac{i}{2} \alpha\, \sigma_{3}\right)\,
\psi, \,\,\, \bar{\psi} \rightarrow \bar{\psi}\,\exp\left(\frac{i}{2}
\alpha\, \sigma_{3}\right), \nonumber\\
\phi &\rightarrow& \exp( -i \alpha)\,\phi,\,\,\,
 \phi^{\dagger} \rightarrow
\exp(i \alpha)\,\phi^{\dagger},\nonumber\\
\lambda &\rightarrow& \lambda - \partial_{\tau} \alpha.
\eq
This gauge invariance comes about because  constraint (3) is time
independent. Using the equations of motion,
$$
\partial_{\tau}\left(\int d\cue\,(\phi^{\dagger} \phi) -\rho\right)=0,
$$
and therefore the constraint is really needed only at a fixed $\tau$,
 say, as an initial condition. This can be implemented by
  gauge fixing by requiring $\lambda$
to be independent of the time $\tau$, 
$$
\lambda(\sigma,\tau)\rightarrow \lambda(\sigma),
$$
by a suitable choice of gauge parameter $\alpha$.  In this time
 independent form,
$\lambda$ is analogous to the chemical potential in statistical
mechanics. After gauge fixing $\lambda$, we have to impose the equation
of motion with respect to it, or, equivalently, the constraint (3)
on the states.

Using the constraint, the free Hamiltonian can be written in a
form more convenient for later application:
\bq
H_{0}&=&\frac{1}{2}\sum_{\sigma,\sigma'}G(\sigma,\sigma')\Bigg(
\frac{1}{2} m^{2}_{0}\,\rho(\sigma) \rho(\sigma') + \rho(\sigma')\,
\int d\cue\,(\cue^{2}+\mu^{2})
\, \phi^{\dagger}(\sigma,\cue) \phi(\sigma,\cue)\nonumber\\
&-&\int d\cue \int d\cue'\,(\cue\cdot \cue')\,
\phi^{\dagger}(\sigma,\cue) \phi(\sigma,\cue)
\phi^{\dagger}(\sigma',\cue') \phi(\sigma',\cue')\Bigg)\nonumber\\
&+&\sum_{\sigma}\lambda(\sigma)\left(\int d\cue\,
 \phi^{\dagger}(\sigma,\cue) \phi(\sigma,\cue) -\rho(\sigma)\right),
\eq
where we have defined
\be
G(\sigma,\sigma')=\frac{\mathcal{E}(\sigma,\sigma')+
\mathcal{E}(\sigma',\sigma)}{|\sigma-\sigma'|}.
\ee
There is a redundancy in the above equation: the mass is split into
two pieces according to
$$
m^{2}=m_{0}^{2}+2 \mu^{2}.
$$
This redundancy will prove useful later on.

Next, we introduce the interaction term. Two kinds of interaction
vertices, corresponding to $\phi^{\dagger}$ creating a solid line
or $\phi$ destroying a solid line, are pictured in Fig.3.
\begin{figure}[t]
\centerline{\epsfig{file=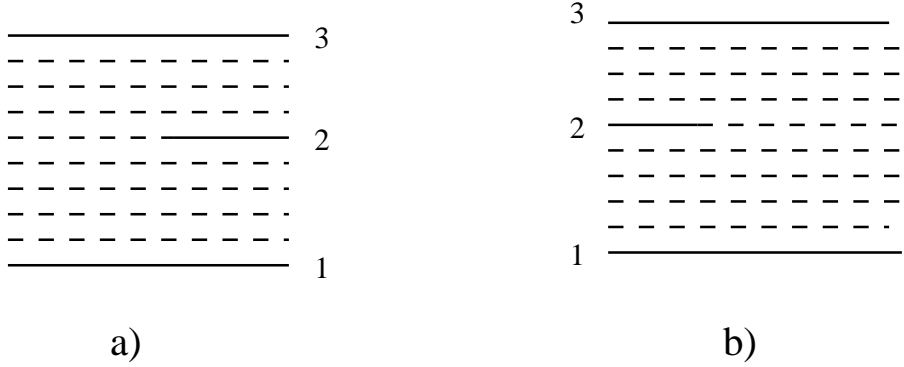, width=12cm}}
\caption{The Two $\phi^{3}$ Vertices}
\end{figure}
  
We take care of the prefactors of the form
 $1/(p^{+})$ in (1) by attaching a factor of
\be
\frac{1}{\sqrt{p_{12}^{+}\, p_{23}^{+}\, p_{13}^{+}}}=
\frac{1}{\sqrt{(\sigma_{2}-\sigma_{1})(\sigma_{3}-\sigma_{2}) 
(\sigma_{3}-\sigma_{1})}}
\ee
to each vertex. The interaction term in the Hamiltonian can
now be written as
\be
H_{I}= g \sqrt{a}\,\sum_{\sigma}\int d\cue\,\left(\mathcal{V}(\sigma)\,
\rho_{+}(\sigma)\, \phi(\sigma,\cue)+
\rho_{-}(\sigma)\,\mathcal{V}(\sigma)\,
\phi^{\dagger}(\sigma,\cue)\right),
\ee
where $g$ is the coupling constant,
\be
\mathcal{V}(\sigma)=\dbsm \frac{W(\sigma_{1},\sigma_{2})}
{\sqsm},
\ee
where,
\be
\wss=\rho(\sigma_{1})\, \mathcal{E}(\sigma_{1},\sigma_{2})\,
 \rho(\sigma_{2}).
\ee
and
\be
\rho_{+}=\bar{\psi}_{1} \psi_{2},\,\,\,\rho_{-}=\bar{\psi}_{2}
\psi_{1}.
\ee

Here is a brief explanation of the origin of various terms in $H_{I}$:
The factors $\rho_{\pm}$
 are there to pair a solid line
with an $i=2$ fermion and a dotted line with an $i=1$ fermion. The
factor of $\mathcal{V}$ ensures that the pair of solid lines 12 and 23
in Fig.3
are separated by only dotted lines, without any intervening solid lines.
Apart from an overall factor, the vertex defined above is very similar
 to the bosonic string interaction vertex in the light cone
picture. Taking advantage of the properties of $\mathcal{E}$ discussed
following eq.(5), we have written an explicit representation of 
this overlap vertex. Finally, the factor of $\sqrt{a}$ multiplying the
 coupling constant is needed to convert sums over $\sigma$ into
integrals in the continuum limit. 

 The total Hamiltonian is given by
\be
H=H_{0}+H_{I}
\ee
and the corresponding action by
\be
S=\int d\tau\left(\sum_{\sigma}\left(i \bar{\psi} \partial_{\tau}
\psi + i\int d\cue\,\phi^{\dagger} \partial_{\tau} \phi \right)
- H(\tau)\right).
\ee

For later use, we note that the theory is invariant under
\be
\phi(\sigma,\tau, \cue)\rightarrow \phi(\sigma,\tau, \cue+{\bf r}),
\ee
where ${\bf r}$ is a constant vector.

\vskip 9pt

\noindent{\bf 4.  Classical Solutions And Mass Renormalization}

\vskip 9pt

In this section,  we look for classical solutions to the
 equations motion resulting from
 the action (16). However, it was
  pointed out in [5]
 that treating the $\rho$'s as classical fields is problematic. It implies
factorization of the expectation values of the products of the $\rho$'s.
For example, 
\be
\langle \rho^{2} \rangle \rightarrow \langle \rho \rangle\,
 \langle \rho \rangle= \rho_{0}^{2},
\ee
and similarly for higher products. 
 On the other hand, treated exactly, $\rho$
takes on only the discrete values $0$ and $1$, and satisfies the
identities
\be
\rho^{2}(\sigma)=\rho(\sigma),\,\,\rho_{+}(\sigma)\rho_{-}(\sigma)= 
1-\rho(\sigma),\,\,\rho_{-}(\sigma)\rho_{+}(\sigma)=\rho(\sigma).
\ee
From these, one can derive two further identities
\bq
G(\sigma,\sigma')\, \rho(\sigma')\,\rho_{-}(\sigma)\,\wss &=&\nonumber\\
\left(\delta_{\sigma',\sigma_{2}}\,\frac{1}{\sigma_{2}- \sigma}+
\delta_{\sigma',\sigma_{1}}\,\frac{1}{\sigma -\sigma_{1}}\right)
&\rho_{-}(\sigma)&\wss,
\eq
and
\be
\wss \rho_{+}(\sigma) \rho_{-}(\sigma) W(\sigma'_{1},\sigma'_{2})
=\delta_{\sigma_{1},\sigma'_{1}}\,\delta_{\sigma_{2},\sigma'_{2}}   
\,\wss.
\ee

 One can also understand eqs.(20, 21) geometrically from the overlap
properties of the vertices in Fig.3.  Apart
from an overall factor, these are structurally the same as the
corresponding string vertices, and in particular, they satisfy the
same overlap relations. These overlap relations turn out to be
crucial for the elimination of both ultraviolet divergences and the
singularity at $a=0$, which is the reflection of the original
$p^{+}=0$ singularity in the propagator
 (1). If present, this singularity would prevent
us from taking the continuum limit of the model.

 The factorization ansatz (18) violates the
identities (19) and consequently the  overlap relations.
To overcome this problem,  we treat
the $\phi$'s as classical fields, but keep the $\rho$'s as operators
satisfying eqs.(19) in the intermediate stages of the computation.
 The  strategy is first to simplify the expressions 
 as much as possible
using the overlap relations (20, 21) before making any approximations.

 We will now
search for solutions $\phi_{0}(\sigma,\cue)$ that are time independent
(solitonic) and whose dependence on $\cue$ is rotationally invariant.
The equation motion for $\phi_{0}$ then simplifies to
\be
\left(2 \lambda(\sigma)+\sum_{\sigma'} G(\sigma,\sigma')\,\rho(\sigma')
\,(\cue^{2}+\mu^{2})\right)\,\phi_{0}(\sigma,\cue) = - 2 g \sqrt{a}\,
\rho_{-}(\sigma)\,\mathcal{V}(\sigma).
\ee

To solve this equation, we make the following ansatz for $\phi_{0}$:
\be
\phi_{0}(\sigma,\cue)=\dbsm\,\rho_{-}(\sigma)\,
 \wss\,\tilde{\phi}_{0}(\sigma,\sigma_{1},\sigma_{2},
\cue).
\ee
where $\tilde{\phi}$ is a c-number and all the operator dependence
is in $\rho_{-} W$. The motivation for this ansatz is that the operators
appearing on either side of the equation are the same; namely
$\rho_{-}(\sigma)\,\wss$. Only the multiplying c-numbers are different.
 To show this for the left hand side of eq.(22), we reduce the operator
 product
$$
G(\sigma,\sigma')\,\rho(\sigma')\wss
$$
 to the form $\rho_{-}(\sigma)\,\wss$ using eq.(20). 
 The right hand side of (22) is also proportional to $\rho_{-}(\sigma)\,\wss$
because of the form of $\mathcal{V}$ (eq.(12)). We can then solve this equation
 by equating the c-number coefficient of the operator $\rho_{-}\,W$
on both sides. The result is
\bq
\tilde{\phi}_{0}(\sigma,\sigma_{1},\sigma_{2},\cue)&=&
 - \frac{2 g\sqrt{a}}{\left(2 \lambda(\sigma)+
(\cue^{2}+\mu^{2})\left(\ssm\right)\right)}\nonumber\\
&\times&\frac{1}{\sqsm},
\eq
and the solution for $\phi_{0}^{\dagger}$ is given by the Hermitian conjugate
expression.

Next, we define $H_{c}$ by replacing $\phi$ by the above $\phi_{0}$
 in the Hamiltonian,
$$
H_{c}=H(\phi=\phi_{0}),
$$
 and simplify again using the overlap relations until
we have a linear result in $W$:
\bq
H_{c}&=& -2 g^{2} a\,\sum_{\sigma} \dbsm\,\int d\cue\, \wss
\nonumber\\
&\times& \left(\qsm\,\left(2 \lambda(\sigma)+
 (\cue^{2} +\mu^{2}) \ssm\right)\right)^{-1}\nonumber\\
&-&\sum_{\sigma} \lambda(\sigma)\,\rho(\sigma)+ \frac{m_{0}^{2}}{2}
\sum_{\sigma'>\sigma}\frac{W(\sigma,\sigma')}{\sigma' -\sigma}.
\eq
Since the overlap relations are quadratic, once we have a linear expression in
$W$, $H_{c}$ cannot be further simplified.

In the above expression, the integral over $\cue$ is ultraviolet divergent
at $D=2$ and $D=4$.
This divergence can be eliminated by the mass renormalization
 and at $D=4$ by also  coupling
constant renormalization. We observe that as $|\cue|\rightarrow \infty$,
the first term on the right, after doing the sum over $\sigma$, reaches
a limit identical in form to the mass term. It can therefore be canceled by
setting
\be
m_{0}^{2}= 4 g^{2} a\,\int d\cue\,\frac{1}{\cue^{2}+\mu^{2}}.
\ee
We note that without the use of the overlap relations, the structure
of the ultraviolet divergence in (25) would be different from the structure
of the mass counter term, and hence renormalization would not be possible.
The presence of $\mu$ avoids an infrared divergence. At $D=2$, there is no
divergence, and at $D=4$, a quadratic divergence is reduced to a logarithmic
divergence in the coupling constant. Although there is no divergence at
$D=1$, we will still use the same expression for $m_{0}$ also in this case.

At the beginning, we started with two independent masses in the
problem. But now  that $m_{0}$ is fixed, only
$\mu$  remains. We could have given a treatment based on a single mass
from the start, however,
having an extra mass temporarily is more convenient. For example, it
 enables us to give a uniform
treatment for all dimensions.

Up to this point, apart from a few changes and amplifications, we have been
reviewing parts of reference [5]. In the next section, we will finally
break new ground. The starting point will be eq.(25). As it stands,
there are two problems with this equation: The world sheet is still
discrete, and the continuum limit $a\rightarrow 0$ is problematic. Also, there
is the operator $W$, which has to be evaluated. Both problems will be
addressed in the next section.

\vskip 9pt

\noindent{\bf 5. The Continuum Limit}

\vskip 9pt

The continuum limit is taken by letting $a\rightarrow 0$, after suitably
scaling the field variables by
\be
\phi\rightarrow \sqrt{a}\,\phi,\,\,\,\psi\rightarrow \sqrt{a}\,\psi,
\ee
and similarly for the hermitian conjugate fields. From its definition,
$\rho$ scales as
\be
\rho\rightarrow a\,\rho.
\ee
 In the limit
$a\rightarrow 0$, the scaled fields satisfy the commutation relations (2),
with, however, the Kroenecker $\delta_{\sigma,\sigma'}$ replaced by the
Dirac  $\delta(\sigma,\sigma')$. From now on, to keep
 things simple, we will continue to
 use the same symbols $\phi$ and $\psi$ also for the 
scaled fields, and whether the discretized or the continuum limit is being
used should be clear from the context.

Let us  now consider the continuum limit of $H_{c}$. In this limit,
all the sigma sums become integrals, and all the factors of $a$ are used up
in this process.
Also, the product in the
definition of $\mathcal{E}$ (eq.(5)) becomes 
\be
\mathcal{E}(\sigma_{1},\sigma_{2})=\prod_{\sigma_{1}}^{\sigma_{2}}
\left(1- a\,\rho(\sigma)\right)
\rightarrow \exp\left(- \int_{\sigma_{1}}
^{\sigma_{2}} d\sigma\,\rho(\sigma)\right).
\ee
 After a change of variables by
$$
\sigma=\sigma_{1}+x\,(\sigma_{2} -\sigma_{1}),
$$
the result can be written as
\bq
H_{c}&=& -2 g^{2} \int d\sigma_{2} \int^{\sigma_{2}} d\sigma_{1}
\int_{0}^{1} dx\int d\cue\, \rho(\sigma_{1})\,
\mathcal{E}(\sigma_{1},\sigma_{2})\,\rho(\sigma_{2})\nonumber\\
&\times &
\left((\sigma_{2} -\sigma_{1})
\left(2 \lambda(\sigma)\,x (1-x)(\sigma_{2} -\sigma_{1})+ 
 (\cue^{2} +\mu^{2})\right)\right)^{-1}\nonumber\\
&+& 2 g^{2}\,\int d\sigma_{2}
\,\int^{\sigma_{2}} d\sigma_{1}
\int d\cue\,\frac{1}{\cue^{2}+\mu^{2}}
\,\frac{\rho(\sigma_{1})\,
\mathcal{E}(\sigma_{1},\sigma_{2})\,\rho(\sigma_{2})}{\sigma_{2} -\sigma_{1}}
- \int d\sigma\, \lambda(\sigma)\,\rho(\sigma).\nonumber\\
&&
\eq

The first and the second terms on the right
 are divergent as $|\cue|\rightarrow \infty$ at
$D=2, 4$, and also they are also logarithmically divergent as
$\sigma_{2}-\sigma_{1}\rightarrow 0$. The first is the ultraviolet mass
 divergence and we have already fixed $m_{0}$ by eq.(26) so that
it cancels between the two terms.
 The second singularity is a logarithmic singularity at
$\sigma_{2}-\sigma_{1}=0$. Since $\sigma_{2}-\sigma_{1}$ is the $p^{+}$
 flowing through the propagator, this is 
 the $p^{+}=0$ singularity in disguise. 
Surprisingly, this divergence also cancels between the first and second terms
 in all dimensions. It is highly satisfying that the mass counter term
introduced to eliminate an ultraviolet divergence also automatically
cancels the infrared divergence at $p^{+}=0$. This cancellation is quite
non-trivial and absolutely essential, since otherwise, having 
only one adjustable constant $m_{0}$ at our disposal, 
 we would be stuck with one divergence or other at  $D=2, 4$. We also
note that we cannot add an arbitrary ultraviolet
 finite term to $m_{0}^{2}$ without spoiling the infrared cancellation.
Although we started with two masses, in the end only $\mu$ remains as
an arbitrary parameter.

Another important feature of $H_{c}$ is its symmetries. In addition to
translation invariance in $\cue$ (eq.(17)),
 as is well known,
the light cone dynamics is manifestly invariant under a subgroup of Lorentz
transformations. The original action (16) is trivially invariant under
under all the generators of this subgroup except for  the generator
$K_{1}$ of boosts along the special direction $1$. The discretization
of the $\sigma$ coordinate breaks this symmetry even at the classical level.
We expect this symmetry will be at least classically
 restored in the continuum limit. To see
this, we note that under $K_{1}$, various fields transform as
\bq
\phi(\sigma,\tau,\cue)&\rightarrow& \sqrt{u}\,\phi(u \sigma, u \tau,\cue),
\,\,\psi(\sigma,\tau,)\rightarrow \sqrt{u}\,\psi(u \sigma, u \tau),\nonumber\\
\rho(\sigma,\tau)&\rightarrow& u\,\rho(u \sigma, u \tau),\,\,
\lambda(\sigma, \tau)\rightarrow u\,\lambda(u \sigma, u \tau),\,\,
p^{+}\rightarrow \frac{1}{u}\,p^{+},
\eq
where $u$ parametrizes the $K_{1}$ transformations. In the expression for
$H_{c}$, this amounts to letting
$$
\sigma\rightarrow u\,\sigma,\,\,\,\tau\rightarrow u\,\tau,
$$
and transforming $\rho$ according to eq.(31). The classical Hamiltonian then
transforms as
\be
H_{c}\rightarrow u\,H_{c},
\ee
and as expected,
the corresponding action is therefore invariant. As we shall see, this
 invariance will be respected by the mean field approximation, and it
will play an important role in what follows.

Eq.(30), which is free of divergences and  independent of $a$, will be
the starting point of the mean field approximation in the next section.
As we stressed earlier, making approximations at an earlier stage could have
easily spoiled these desirable features.

\vskip 9pt

\noindent{\bf 6. The Meanfield Approximation}

\vskip 9pt

The mean field approximation consists of replacing $\rho$ and $\lambda$
in $H_{c}$ by their ground state expectation values, which we assume to be
independent of $\sigma$ and $\tau$.
(translation invariance of the ground state).
 Afterwards,  $H_{c}$ should be minimized, subject to the constraints
(3) and a (gauge) fixed  $\lambda$, to
find the ground state. Eq.(30) can then be simplified by the following change 
of variables:
\be
\lambda= \lambda_{0}\,\rho,\,\,\,\sigma= \sigma'/\rho.
\ee
In terms of the new variables, we have,
\bq
H_{c}&=& p^{+}\,\rho^{2}\Bigg(- \lambda_{0}+ 2 g^{2}\,\int_{0}^{\rho\,p^{+}}
d\sigma' \int d\cue\,\left(\frac{\exp(-\sigma')}{\sigma'}\right)\nonumber\\
&\times&\left(\frac{1}{\cue^{2}+\mu^{2}}-
 \frac{1}{\cue^{2}+\mu^{2}+ 2 \lambda_{0}
\,x (1-x) \sigma'}\right)\Bigg).
\eq

Let us now examine the structure of this equation. It can be written as
\be
H_{c}= p^{+}\,\rho^{2}\,F(\lambda_{0},\,\rho\,p^{+})
\ee
The dependence of $F$ on $\rho\,p^{+}$ comes solely from the upper
limit in the $\sigma'$ integration. One could study $H_{c}$
as a function of $\rho$ and $\lambda_{0}$ without making any further
approximations other than mean field, but we shall not attempt it in this
paper. Instead, we will
exclusively study the model in the small $\lambda_{0}$ limit. One motivation
for this limit is that it is possible to get simple explicit results. 
 Also, having
already fixed $\lambda_{0}\,p^{+}$ (see the discussion following
eq.(7)), this corresponds to taking $\rho\,p^{+}$ large. Since  $\rho\,p^{+}$
counts the total number of solid lines (propagators) on the world sheet, this
corresponds to a world sheet densely covered
 with a large number of solid lines (propagators). This limit is of
particular interest, since it is clearly non-perturbative and difficult to
study by other methods. Also, it is in this limit that we may
expect string formation on the world sheet.

From eq.(34), we see that in the large  $\rho\,p^{+}$ limit, we can let
 $\rho\,p^{+}\rightarrow \infty$, and suppress the explicit dependence on
 $\rho\,p^{+}$ in $F$.
 This amounts to
neglecting exponentially suppressed
 boundary effects coming from the finite range 
of the $\sigma'$ coordinate. Notice, however, that it is $\rho\,p^{+}$
and not $p^{+}$ that is taken large. Letting $p^{+}$ become large is not
$K_{1}$ invariant condition, since $p^{+}$ scales under $K_{1}$. In 
contrast,  $\lambda_{0}\,p^{+}$ is $K_{1}$ invariant and taking it
large is frame independent.

After dropping the dependence on the boundary effects,
we can understand the structure of this equation
 as follows: The single factor of $p^{+}$
reflects the fact that on the world sheet that is uniform in $\sigma$,
 $H_{c}$ is extensive in  $p^{+}$. The factor of
$\rho^{2}$ is needed to have $H_{c}$ transform correctly under $K_{1}$
(see eq.(31)). The function $F$ could be arbitrary as far as $K_{1}$
invariance is concerned, since $\lambda_{0}=\lambda/\rho$ is $K_{1}$
invariant.  The general argument
 we have given shows that  eq.(35) is valid
 in all transverse dimensions $D$ and even
after the addition of a $\phi^{4}$ interaction.

 We will determine $F$ through the  equation of motion with respect to
$\lambda$, which, at fixed $\rho$, is the same as the equation of motion with
respect to $\lambda_{0}$:
\be
\frac{\partial H_{c}}{\partial \lambda_{0}}=0 \rightarrow
\frac{\partial F}{\partial \lambda_{0}}=0.
\ee
Since $\lambda$ is fixed as an initial condition, this amounts to
the determination of $\rho$ in terms of $\lambda$.
 Carrying out the integration over $\cue$ in various dimensions, we
have
\be
F_{D}= -\lambda_{0}+C_{D}\, g^{2}\,\int_{0}^{\infty} d\sigma'
\int_{0}^{1} d x\, \frac{\exp(-\sigma')}{\sigma'}\
 L_{D}(x,\sigma',\lambda_{0}),
\ee
where,
\be
C_{1}= 2\pi,\,\,C_{2}= 2\pi,\,\,C_{4}=2 \pi^{2},
\ee
and,
\bq
 L_{1}&=&\frac{1}{\mu} -
\frac{1}{\left(\mu^{2}+ 2\lambda_{0}\,x (1-x)
\sigma'\right)^{1/2}},\nonumber\\
 L_{2}&=&\ln\left(1+\frac{2\lambda_{0}\,x (1-x)\,\sigma'}
{\mu^{2}}\right),\nonumber\\
L_{4}&=&2\lambda_{0}\,x (1-x)\,\sigma'\,\ln\left(\frac{\Lambda^{2}}
{\mu^{2}}\right)\nonumber\\
&-&\mu^{2}\left(1+\frac{2\lambda_{0}\,x (1-x)\,\sigma'}
{\mu^{2}}\right)\,\ln\left(1+\frac{2\lambda_{0}\,x (1-x)\,\sigma'}
{\mu^{2}}\right).
\eq

In the last equation, $\Lambda$ is an ultraviolet cutoff. These equations
fix the dimensionless parameter $\lambda_{0}/\mu^{2}$ in terms of the
dimensionless constants $g^{2}/\mu^{3}$ at $D=1$ and $g^{2}/\mu^{2}$
at $D=2$. At $D=4$, the relation between $g^{2}$ and $\lambda_{0}/\mu^{2}$
is more complicated and it is discussed below. We will explicitly
evaluate these relations only in the small $\lambda_{0}$ limit by
expanding to first order in  $\lambda_{0}$ (see the discussion following
eq.(35)):
\bq
\frac{\partial F_{1}}{\partial \lambda_{0}}&=&0\rightarrow
\frac{\lambda_{0}}{\mu^{2}}= \frac{5}{3}- \frac{5 \mu^{3}}{\pi\,g^{2}}
+\cdots,\nonumber\\
\frac{\partial F_{2}}{\partial \lambda_{0}}&=&0\rightarrow
\frac{\lambda_{0}}{\mu^{2}}=\frac{5}{2}- \frac{15 \mu^{2}}{4 \pi\,g^{2}}
+\cdots,\nonumber\\
\frac{\partial F_{4}}{\partial \lambda_{0}}&=&0\rightarrow
\frac{1}{g^{2}}=\pi^{2}\left(\frac{2}{3}\left(\ln(\Lambda^{2}/\mu^{2})
-1\right) -\frac{4 \lambda_{0}}{15 \mu^{2}}+\cdots\right).
\eq

The first two equations give us conditions on the coupling constant.
Since the left hand sides of these equations are positive, it
follows that
$$
\frac{g^{2}}{\mu^{3}}\geq \frac{3}{\pi}
$$
at $D=1$ and,
$$
\frac{g^{2}}{\mu^{2}}\geq \frac{3}{2 \pi}
$$
at $D=2$. Otherwise, there is no solution to these equations. At least in
this approximation, these inequalities are then the conditions for the
formation of a world sheet densely populated with solid lines.
We note that expanding
 in $\lambda_{0}$, is the same as expanding in the small
parameter
$$
1- \frac{3 \mu^{3}}{\pi g^{2}}
$$
at $D=1$ and in 
$$
1-\frac{3 \mu^{2}}{2 \pi g^{2}}
$$
at $D=2$. Although we will not pursue it here, one can envisage a
systematic expansion in these small parameters.

 In the case of $D=4$,
$\lambda_{0}$ need not really be small,
$$
\Lambda^{2}/\mu^{2}\gg 1,
$$
is all that is needed, as discussed below. In this case, the expansion
parameter is the running coupling constant.

The last equation, at $D=4$, has a logarithmic dependence on the
cutoff $\Lambda$. This is related to the coupling constant renormalization.
 We recall that $\phi^{3}$ is asymptotically free
in 6 space-time dimensions ($D=4$),
 and the above relation is the well known lowest order renormalization
group result obtained by summing the leading logarithmic
divergences in the perturbation series. To get a finite result,
 one should first
renormalize the coupling constant before summing the logs. This
amounts to replacing the cutoff $\Lambda$ by a large but finite
value. The coupling constant on the left should then be identified
with  the running coupling constant $g(\Lambda)$, defined at the energy
scale  $\Lambda$. For this leading log. approximation to be reliable,
$g(\Lambda)$ should be small, which means that $\Lambda^{2}/\mu^{2}$
should be large. All the additional terms on the right hand side,
including the term proportional to $\lambda_{0}$, only
make a small change in the scale
of the running coupling constant.

An interesting quantity is the ground state energy. It can be
computed to lowest order in $\lambda$
from equations (35), (37) and (40):
\bq
H_{c}(D=1)&=& \frac{3}{10}\,\frac{p^{+}\,\lambda^{2}}{\mu^{2}}+\cdots
=\frac{3}{10}\,\frac{\mu^{2}\,\bar{\lambda}^{2}}{p^{+}}+\cdots,
\nonumber\\
H_{c}(D=2)&=& \frac{3}{10}\,\frac{p^{+}\,\lambda^{2}}{\mu^{2}}+\cdots
=\frac{3}{10}\,\frac{\mu^{2}\,\bar{\lambda}^{2}}{p^{+}}+\cdots,
\nonumber\\
H_{c}(D=4)&=&\frac{1}{5\left(\ln\left(\Lambda^{2}/\mu^{2}\right)-1
\right)}\,\frac{p^{+}\,\lambda^{2}}{\mu^{2}}+\cdots=
\frac{1}{5\left(\ln\left(\Lambda^{2}/\mu^{2}\right)-1
\right)}\,\frac{\mu^{2}\,\bar{\lambda}^{2}}{p^{+}}+\cdots,\nonumber\\
&&
\eq
where,
$$
\bar{\lambda}=\frac{p^{+}\,\lambda}{\mu^{2}}
$$
is a pure number; it is dimensionless and invariant under $K_{1}$
(see eq.(31)).  The light cone energy is $p^{-}$, so multiplying it
by $p^{+}$, we find that the invariant mass squared of
the ground state is proportional to $\bar{\lambda}^{2}\,\mu^{2}$.
 Notice that the mass squares are all positive.
At least, in this approximation, there is no instability.

\vskip 9pt

\noindent{\bf 7. Singularity Cancellation: $\phi^{3}+\phi^{4}$
 In 1+2 Dimensions}

\vskip 9pt

In this section, we add a $\phi^{4}$ interaction to the $\phi^{3}$
of the previous section, and show that, just as in the
case of pure $\phi^{3}$,
 the singularity at $p^{+}=0$ can be canceled by a mass counter
term. The idea is to demonstrate that the elimination of this singularity
is not a specific feature of the $\phi^{3}$ interaction but it can be
generalized to include an additional $\phi^{4}$ term. We will consider
the model only at $D=1$, and thereby avoid the complications resulting
from the renormalization of the $\phi^{4}$ coupling constant at $D=2$.
These complications can be handled just as in the case of $\phi^{3}$
at $D=4$, but we will skip it in the interests of simplicity
and brevity. We will also not study the model in any detail, beyond
showing the elimination of the singularity.

 The total interaction Hamiltonian is with this addition
is given by
\be
H_{I,t}= H_{I}+ H'_{I},
\ee
where $H_{I}$ is given by eq.(11) and,
\bq
H'_{I}&=& g'\,a\,\sum_{\sigma_{1,2,3,4}}\,
\int d \cue \int d \cue'
\,\phi^{\dagger}(\sigma_{2}, \cue)\,\rho_{-}(\sigma_{2})\,
\mathcal{E}(\sigma_{1},\sigma_{4})\,\rho_{+}(\sigma_{3})\,
 \phi(\sigma_{3}, \cue')\nonumber\\
&\times&\left((\sigma_{2} -\sigma_{1})\,(\sigma_{4} -\sigma_{2})\,
(\sigma_{4} -\sigma_{3})\,(\sigma_{3} -\sigma_{1})\right)^{- 1/2}
+\cdots.
\eq
Here, $g'$ is a positive coupling constant, scaled by $a$ in order
that in the limit $a\rightarrow 0$, sums over $\sigma$ smoothly go
over integrals over $\sigma$ (see section 3). The sum over the
$\sigma$'s is non-zero only for
$$
\sigma_{1}<\sigma_{2}<\sigma_{4},\,\,\,\sigma_{1}<\sigma_{3}<\sigma_{4}.
$$

 Actually, the term we have written out in
the expression for $H'_{I}$ corresponds to first graph in Fig.4 .
\begin{figure}
\centerline{\epsfig {file=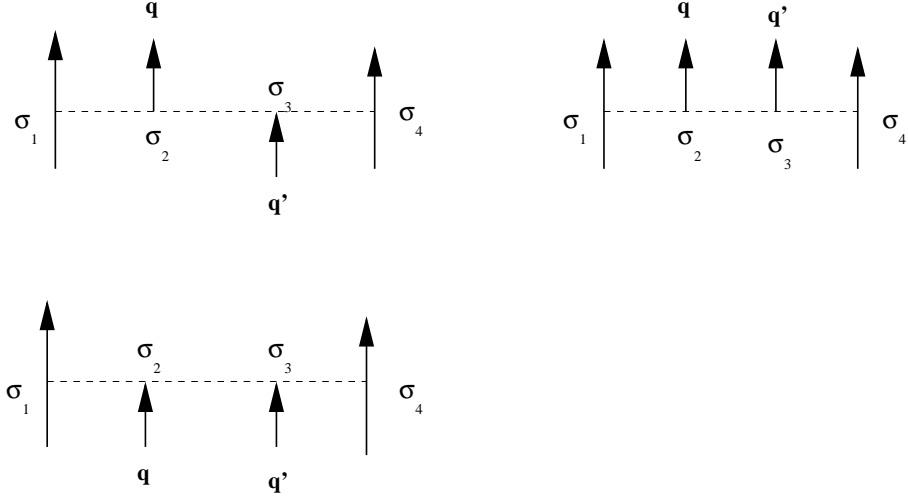, width= 12cm}}
\caption{Four Point Vertices}
\end{figure}
 We will demonstrate the elimination
of the singularity only for this graph; the other two graphs can be
treated along the same lines.

We are now ready to write down  the equations of motion,
similar to (22), for this extended model. Actually, the algebra can be 
greatly simplified by setting $\lambda=0$ at the beginning. This is because
singularity cancellation occurs at $\lambda=0$, or equivalently, at
$\lambda_{0}=0$. In the case of $\phi^{3}$ interaction, this can be seen
from eq.(25): Expanding the integral on the right hand side in powers of
$\lambda_{0}=0$, we notice that the zeroth order terms cancel, and 
 the higher order terms depend only on the powers of $\lambda_{0}=0$
 in the combination
$\lambda_{0}\,\sigma'$. Therefore, the potential singularity at
$\sigma'=0$ is absent because of the extra factors of $\sigma'$.
This is true even after the addition of a $\phi^{4}$ interaction;
higher order terms in $\lambda_{0}$ always come with extra powers of
$\sigma'$ and therefore they are singularity free.

The equation of motion for $\phi_{0}(\sigma)$,
\bq
&&\sum_{\sigma'} G(\sigma,\sigma')\,\rho(\sigma')
\,(\cue^{2}+\mu^{2})\,\phi_{0}(\sigma,\cue) = 2 g \sqrt{a}\,
\rho_{-}(\sigma)\,\mathcal{V}(\sigma)\nonumber\\
&+&g' a\int d \cue'\sum_{\sigma_{1}<\sigma'<\sigma_{2}}\frac{
\rho_{-}(\sigma)\,W(\sigma_{1},\sigma_{2})\,\phi_{0}(\sigma',\cue')}
{\left((\sigma_{2}-\sigma) (\sigma -\sigma_{1}) (\sigma_{2}-\sigma')
 (\sigma' -\sigma_{1})\right)^{1/2}},
\eq
differs from eq.(22) in two respects: $\lambda$ is set equal to zero
and there is an extra term on the right hand side coming from the
$\phi^{4}$ interaction. It turns out that this equation is solved
by the  ansatz 
\be
\phi_{0}(\sigma,\cue)=2 A \sqrt{a}\,\sum_{\sigma_{1}<\sigma<\sigma_{2}}
\frac{\rho_{-}(\sigma)\,W(\sigma_{1},\sigma_{2})\,\left((\sigma-\sigma_{1})
(\sigma_{2}-\sigma)\right)^{1/2}}{(\cue^{2}+\mu^{2})\,(\sigma_{2}-\sigma_{1})
^{3/2}}.
\ee
This ansatz is almost
 the same as eq.(23), with the only difference
that $\lambda=0$ and the overall
coefficient $A$ is left arbitrary. Substituting in the equation of motion (44)
and simplifying by the use of the overlap relations, we find that if
\be
A=-\left(g+\frac{2 \pi}{\mu}\,A\,g'\right),
\ee
the equation of motion is satisfied. The resulting $H_{c}$ has the same form
as eq.(34), only with $\lambda_{0}=0$ and $g^{2}$ replaced
by $A^{2}$. The mass counter term $m_{0}^{2}$ is  given by eq.(26) as before,
with again  $g^{2}$ replaced by  $A^{2}$. With these changes,
just as in the case of pure $\phi^{3}$, the
result is still non-singular at $\sigma'=0$.
Therefore, even with the addition of a
$\phi^{4}$ interaction, one can  go to
the continuum limit without encountering any singularities.

\vskip 9pt

\noindent{\bf 8.  Fluctuations Around The Classical
Background} 

\vskip 9pt

Given the classical solutions developed in the previous sections, it is
natural  to study quantum fluctuations about these
backgrounds. This can be done explicitly to  quadratic order for all the
 fluctuations. We will, instead, focus exclusively on a particular
set of fluctuations, which were studied also in the earlier work . These
are obtained by quantizing the collective coordinates corresponding
to broken translation invariance
\be
\cue\rightarrow \cue+ {\bf r}.
\ee
(See eq.(17)). The classical solution, placed at a definite
location in the $\cue$ space, breaks this symmetry. This is familiar from
solitons in field theory, and 
the symmetry is restored by quantizing the so-called collective modes.
These modes
are very important not only
for their role in restoring translation
invariance, but also, because, they are the low lying Goldstone modes
connected with the spontaneously broken translation symmetry.
In the earlier work, they were crucial to the
formation  of a string on the world sheet. Also, as
we shall see, their contribution to the action can be computed exactly.
For all these reasons, we will study only these collective modes and 
 not consider any other 
modes in this work.

Inevitably, there will be some overlap with the earlier work [5].
However, we are not here trying merely to reproduce previous results;
 in fact, with our new approach, we arrive at a somewhat different
picture, such as no-linear string trajectories, which we will explain shortly.

The collective coordinate corresponding to translations is
introduced by letting
\be
\phi=\phi_{0}+\phi_{1},
\ee
where $\phi_{1}$ is the fluctuating part of the field, and setting,
\be
\phi_{1}(\sigma,\tau,\cue)=
 \phi_{0}(\sigma,\cue+\bfv(\sigma,\tau))
-\phi_{0}(\sigma,\cue),
\ee
where $\phi_{0}$ is the classical solution and
$\bfv$ is the collective coordinate. The contribution
of $\phi_{1}$ to the action
can be written as the sum of kinetic and potential terms:
\be
S^{(1)}= S_{k.e}- \int d\tau\,H_{0}(\phi_{1})=S_{k.e}+ S_{p.e},
\ee
where the kinetic term depends on $\partial_{\tau}\bfv$  and the potential
has no $\tau$ derivatives. We note that only $H_{0}$ contributes
 to $S_{p.e}$:
There are no $\tau$ derivatives and the linear  $\phi$ terms in
$H_{I}$ have been eliminated  by shifting $\phi$ by the classical solution.

We now substitute the ansatz (49) directly into $H_{0}$ (eq.(8)).
The integrals over $\cue$ and $\cue'$ can be done
 by shifting them by $\bfv$, and the result can be simplified by invoking
the constraint (3) and the rotation invariance of $\phi_{0}$, with the
result:
\be
 S_{p.e}= -\frac{1}{4}\int d\tau \int d\sigma \int d\sigma'\,
\frac{W(\sigma.\sigma')}{|\sigma-\sigma'|}\left(\bfv(\sigma,\tau)
- \bfv(\sigma',\tau)\right)^{2}.
\ee

We note that so far no approximation was made, and therefore, this
result is exact so long as only the contribution of the collective
coordinate $\bfv$ is concerned. Also,
 there is no singularity at $\sigma=\sigma'$ and so there
is no obstacle to taking the continuum limit immediately. To make
further progress, we introduce the
mean field approximation by setting  $\rho$ to be a constant, with the
result
\be
 S_{p.e}\rightarrow -\frac{\rho^{2}}{4} \int d\tau \int d\sigma 
\int d\sigma'\, \frac{\exp(-\rho\,|\sigma-\sigma'|)}{|\sigma-\sigma'|}
\left(\bfv(\sigma,\tau)- \bfv(\sigma',\tau)\right)^{2}.
\ee

We will study this action in detail later on, but before that, we turn
our attention to the kinetic energy term. We will first compute this
term to quadratic order in $\partial_{\tau}\bfv$, and later argue that
the result is exact. Since the action (16) is first order in the $\tau$
derivative, to cast it into a quadratic form in $\partial_{\tau}\phi_{1}$,
 one has to split $\phi_{1}$
into its real and imaginary (Hermitian and anti-Hermitian) parts:
\be
\phi_{1}= \phi_{1,r}+\phi_{1,i},
\ee
and eliminate one of them by integrating over it. In this case, since
the classical solution $\phi_{0}$ is real,  the ansatz (49) more
precisely applies only
to the real part of $\phi_{1}$:
\be
 \phi_{1,r}(\sigma,\tau,\cue)\rightarrow
 \phi_{0}(\sigma,\cue+\bfv(\sigma,\tau)),
\ee
and $\phi_{1,i}$ will be integrated out. The kinetic energy term in the
action (16) can then be rewritten as
\bq
&& i \sum_{\sigma} \int d\tau
\int d\cue\,\phi^{\dagger} \partial_{\tau} \phi= 2 \sum_{\sigma}
 \int d\tau \int d\cue\,\phi_{1,i}\,\partial_{\tau}\phi_{1,r}
\nonumber\\
&\rightarrow& 2 \sum_{\sigma}\int d\tau 
\int d\cue\,\phi_{1,i}\,\partial_{\tau}
 \phi_{0}(\sigma,\cue+\bfv(\sigma,\tau)).
\eq

Now consider the contribution coming from $H_{0}$. We have already taken
care of the contribution of $\phi_{1,r}$ when we computed $S_{p.e}$, so
it remains to compute the quadratic terms in $\phi_{1,r}$. Integrating 
over  $\phi_{1,i}$ then amounts to solving the equations of motion for
 $\phi_{1,i}$ and substituting in the action. The left hand side of the
equation of motion is the same as in (22), but the right hand side comes from
the  variation of the above kinetic term  with respect to  $\phi_{1,i}$:
\be
\left(2 \lambda(\sigma)+\sum_{\sigma'} G(\sigma,\sigma')\,\rho(\sigma')
\,(\cue^{2}+\mu^{2})\right)\,\phi_{1,i}(\sigma,\tau,\cue)=
2\partial_{\tau}\phi_{0}(\sigma,\cue+\bfv(\sigma,\tau)).
\ee
This equation can be solved by letting
\be
\phi_{1,i}(\sigma,\tau,\cue)=\dbsm\,\rho_{-}(\sigma)\,
 \wss\,\tilde{\phi}_{1.i}(\sigma,\sigma_{1},\sigma_{2},\tau,
\cue),
\ee
as in (23).
Following the same steps  as before, this can then be simplified using
the overlap relations, and after some algebra, we have the solution
\be
\tilde{\phi}_{1.i}(\sigma,\sigma_{1},\sigma_{2},\tau)=
\frac{2\, \partial_{\tau}\bfv(\sigma.\tau)\cdot{\bf \bigtriangledown}_{q}
\tilde{\phi}_{0}(\sigma,\sigma_{1},\sigma_{2},\tau,\cue)}
{\left(2\, \lambda(\sigma)+
(\cue^{2}+\mu^{2})\left(\ssm\right)\right)},
\ee
where $\tilde{\phi}_{0}$ is given by (24). Substituting this  back
in the action, and after some more use of the overlap relations, we arrive
at a result free of singularities. We can then take the continuum limit, and
apply the mean field approximation. We skip the intermediate steps of the
straightforward algebra and give the final result:
\be
S_{k.e}\rightarrow\int d\tau \int d\sigma\, \frac{1}{2}\,E(\lambda_{0})\,
\left(\partial_{\tau} \bfv(\sigma,\tau)\right)^{2},
\ee
where,
\be
E(\lambda_{0})=\int_{0}^{\infty} d \sigma_{1} \int_{0}^{\infty} d \sigma_{2}
\int d\cue\,\frac{ 64\,g^{2}\,\cue^{2}}{D}\,
\frac{\sigma_{1}^{2}\,\sigma_{2}^{2}
\,(\sigma_{1} +\sigma_{2})\,\exp\left(-(\sigma_{1} +\sigma_{2})\right)}
{\left(2\,\lambda_{0}\,\sigma_{1}\,\sigma_{2} +(\cue^{2}+\mu^{2})\,
(\sigma_{1} +\sigma_{2})\right)^{5}}.
\ee

A few comments about this result are in order:\\
a) The integrals in the expression for E are convergent in the dimensions
we are considering, and therefore, there are no problems with singularities.\\
b) E is independent of $\sigma$. This follows quite generally from 
translation invariance in the $\sigma$ coordinate in the mean field
approximation. As a consequence, $S_{k.e}$ is local both in $\sigma$
and $\tau$. In contrast, $S_{p.e}$ is non-local in $\sigma$ (eq.(52)).
Since $\lambda_{0}$ is already determined in terms of $g$ and $\mu$
(see eq.(40)),  E is just a fixed normalization constant.

From the derivation given above, it may seem that we have  only
 taken into account the quadratic terms in $\phi_{1}$.
However, we shall now argue that the result for $S_{k.e}$
 is more general, and does not depend on this approximation. This follows
 from the following properties of  $S_{k.e}$: It is local, it is rotation
invariant in $\cue$, and it invariant under 
 \be
\bfv(\sigma,\tau)\rightarrow \bfv(\sigma,\tau)+ {\bf r},
\ee
 ${\bf r}$ is a constant vector. This follows from the translation invariance
of $\cue$ (eq.(17).
A general term local, second order in $\partial_{\tau}$, and rotation 
invariant has to be of the form
$$
\Delta S= \int d\tau \int d\sigma \int d\cue\,\left(\partial_{\tau}
\bfv(\sigma,\tau)\right)^{2}\,Z\left(\left(\bfv(\sigma,\tau)\right)^{2}
\right),
$$
but this is not invariant under the translation of $\bfv$ by ${\bf r}$ unless
the function $Z$ is a
constant. The remaining possibility is a term fourth order in 
$\partial_{\tau}$, of the form
$$
\left(\left(\partial_{\tau}
\bfv(\sigma,\tau)\right)^{2}\right)^{2},
$$
 coming from the fourth order term 
$$
- \frac{1}{2}\sum_{\sigma,\sigma'}\,G(\sigma,\sigma')
\int d\cue \int d\cue'\,(\cue\cdot \cue')\,
\phi^{\dagger} \phi(\sigma,\cue)\,
\phi^{\dagger} \phi(\sigma',\cue')
$$
in $H_{0}$,
but rotation invariance in $\cue$ forbids such a term. Of course, this simple
result holds only for the contribution coming from $\bfv$; all sorts of
other quantum fluctuations have been neglected.

\vskip 9pt

\noindent{\bf 9. String Formation}

\vskip 9pt

In this section, we will study the spectrum of the collective coordinate
$\bfv$, with the action  given by the
sum of $S_{p.e}$ (eq.(52)) and  $S_{k.e}$ (eq.(59)). This is a free
field theory and therefore it is exactly solvable. The kinetic energy is
 already diagonal, and the potential term can be diagonalized by defining
$$
\bfv(\sigma)=\frac{1}{2 \pi}\int dl\,e^{-i l \sigma}\,\tilde{v}(l).
$$
In terms of $\tilde{v}$, the Hamiltonian $H_{p.e}$ corresponding to
  $S_{p.e}$ becomes diagonal:
\be
 H_{p.e} =- \frac{\rho^{2}}{4 \pi}\,\int dl\,\tilde{v}(l)\cdot
\tilde{v}(-l)\,\ln\left(\frac{\rho^{2}}{\rho^{2}+l^{2}}\right).
\ee
So far, we have been treating $l$ as a continuous variable, but since 
$\sigma$ is compactified on circle of circumference $p^{+}$, $l$
should be discretized:
$$
l\rightarrow  \frac{2 \pi\,n}{p^{+}}.
$$
The integral over $l$ is then replaced by
$$
\int dl\rightarrow \frac{2 \pi}{p^{+}}\,\sum_{l}.
$$

Now consider the limit of large $\rho$, which corresponds to 
 $\lambda_{0}\rightarrow 0$. More precisely, we consider the range
$l\ll \rho$.\ so that
$$
\ln\left(\frac{\rho^{2}}{\rho^{2}+l^{2}}\right)\rightarrow
- \frac{l^{2}}{\rho^{2}},
$$
to the lowest order in $l$, and,
\be
 H_{p.e}\rightarrow \frac{2}{p^{+}}\,\sum_{l} l^{2}\,\tilde{v}(l)\cdot
\tilde{v}(-l)= \frac{1}{2}\int_{0}^{p^{+}} d\sigma\left(\partial_{\sigma}
\bfv (\sigma)\right)^{2}.
\ee
In this limit, the action for $\bfv$ tends to the string action:
\be
S\rightarrow \frac{1}{2}\,\int d\tau \int_{0}^{p^{+}} d\sigma
\left( E\,\left(\partial_{\tau} \bfv(\sigma,\tau)\right)^{2}
-\left(\partial_{\sigma} \bfv(\sigma,\tau)\right)^{2}\right).
\ee
However, this is true only in extreme limit $\rho\rightarrow \infty$.
For a finite $\rho$, no matter how large, the Regge trajectories are
no longer linear, and therefore, the string picture is only
approximately valid for  relatively low lying states. Noting that
we are dealing with a scalar field theory, this deviation from the
strict string picture should not be surprising.

\vskip 9pt

\noindent{\bf 10. Conclusions}

\vskip 9pt

As we have already mentioned in the introduction, we expect the present
work to conclude a long series of investigations of scalar field theory
on the light cone world sheet. We have here used the methods and
tools developed in especially reference [5]. Where we differ from this
reference is in the handling of the continuum limit $a\rightarrow 0$.
We show that with an appropriate choice of the mass counter term, both
the singularity at $a=0$  and the ultraviolet mass divergences
are eliminated. This is an important advance; we have finally
a result free of cutoff, which made the previous work unreliable. The
results about the ground state of the model and string formation
do not differ drastically from the earlier results, but they are now
much more trustworthy.

There remains still the problem of manifest Lorentz invariance. Within the
framework of the approach used here, an initial step in this direction
has been taken [13]. It remains to see how far one can push it as a
practical method. Also, elimination of the cutoff dependence makes it easier
to generalize to more interesting theories, such as gauge theories. One of the
simplest of these, non-Abelian theory in 1+2 dimensions, seems to be almost
 within reach.

\vskip 9pt

\noindent{\bf Acknowledgment}

\vskip 9pt

This work was supported in part by the director, Office of Science,
Office of High Energy Physics of the U.S. Department of Energy under Contract
DE-AC02--05CH11231.

\vskip 9pt

\noindent{\bf References}

\vskip 9pt

\begin{enumerate}

\item K.Bardakci and C.B.Thorn, Nucl.Phys. {\bf B 626} (2002)
287, hep-th/0110301.
\item G.'t Hooft, Nucl.Phys. {\bf B 72} (1974) 461.
\item C.B.Thorn, Nucl.Phys. {\bf B 637} (2002) 272, hep-th/0203167,
S.Gudmundsson, C.B.Thorn and T.A.Tran, Nucl.Phys. {\bf B 649} 92003)
3-38, hep-th/0209102.
\item C.B.Thorn and T.A.Tran, Nucl.Phys. {\bf B 677} (2004) 289,
hep-th/0307203.
\item K.Bardakci, JHEP {\bf 1110} (2011) 071, arXiv:1107.5324.
\item C.B.Thorn, Phys.Rev. {\bf D 82} (2010), arXiv:1010.5998.
\item H.P.Nielsen and P.Olesen, Phys.Lett. {\bf B 32} (1970) 203.
\item B.Sakita and M.A.Virasoro, Phys.Rev.Lett. {\bf 24} (1970) 1146.
\item K.Bardakci, JHEP {\bf 0810} (2008) 056, arXiv:0808.2959.
\item A.Casher, Phys.Rev. {\bf D 14} (1976) 452.
\item R.Giles and C.B.Thorn, Phys.Rev. {\bf D 16} (1977) 366.
\item C.B.Thorn, Nucl.Phys. {\bf B 699} (2004) 427, hep-th/0405018,
D.Chakrabarti, J.Qiu and C.B.Thorn, Phys.Rev. {\bf D 74} (2006)
045018, hep-th/0602026.
\item K.Bardakci, JHEP {\bf 1207} (2012) 179, arXiv:1206.1075. 
\item K.Bardakci, JHEP {\bf 0903} (2009) 088, arXiv:0901.0949.

\end{enumerate}

\end{document}